\documentclass[aps,prd,twocolumn,eqsecnum,showpacs,amsmath,nofootinbib]{revtex4}%

\usepackage[dvips]{color,graphicx}
\usepackage{amsfonts,amssymb,theorem,mathrsfs,times}
\newcommand{\ma}[1]{\mbox{$\mathcal{#1}$}}
\newcommand{\qed}{\hbox{\rule[-2pt]{6pt}{6pt}}}
\newcommand{\D}{{\rm d}}

{\theorembodyfont{\upshape}
\newtheorem{Prop}{Proposition}}
{\theorembodyfont{\upshape}
}
{\theorembodyfont{\upshape}
\newtheorem{lm}{Lemma}}
{\theorembodyfont{\upshape}
\newtheorem{dn}{Definition}}
{\theorembodyfont{\upshape}
}
{\theorembodyfont{\upshape}
\newtheorem{Conj}{Conjecture}}

\newcommand{\dalm}{\kern1pt\vbox{\hrule height 0.9pt\hbox{\vrule width
0.9pt\hskip 2.5pt\vbox{\vskip 5.5pt}\hskip 3pt\vrule width 0.3pt}\hrule height
0.3pt}\kern1pt}

\begin{document}

\title{
Generalized Misner-Sharp quasi-local mass in Einstein-Gauss-Bonnet gravity
}

\author{Hideki Maeda$^{1,2}$}
\email{hideki@cecs.cl}
\author{Masato Nozawa$^3$}
\email{nozawa@gravity.phys.waseda.ac.jp}


\address{ 
$^{1}$Centro de Estudios Cient\'{\i}ficos (CECS), Arturo Prat 514, Valdivia, Chile\\
$^{2}$Department of Physics, International Christian University, 
3-10-2 Osawa, Mitaka-shi, Tokyo 181-8585, Japan\\
$^{3}$Department of Physics, 
Waseda University, Tokyo 169-8555, Japan
}

\date{\today}

\begin{abstract} 
We investigate properties of a quasi-local mass in a higher-dimensional spacetime having symmetries corresponding to the isomertries of an $(n-2)$-dimensional maximally symmetric space in Einstein-Gauss-Bonnet gravity in the presence of a cosmological constant.
We assume that the Gauss-Bonnet coupling constant is non-negative.
The quasi-local mass was recently defined by one of the authors as a counterpart of the Misner-Sharp quasi-local mass in general relativity.
The quasi-local mass is found to be a quasi-local conserved charge associated with a locally conserved current constructed from the generalized Kodama vector and exhibits the unified first law corresponding to the energy-balance law.
In the asymptotically flat case, it converges to the Arnowitt-Deser-Misner mass at spacelike infinity, while it does to the Deser-Tekin and Padilla mass at infinity in the case of asymptotically AdS.
Under the dominant energy condition, we show the monotonicity of the quasi-local mass for any $k$, while the positivity on an untrapped hypersurface with a regular center is shown for $k=1$ and for $k=0$ with an additional condition, where $k=\pm1,0$ is the constant sectional curvature of each spatial section of equipotential surfaces.
Under a special relation between coupling constants, positivity of the quasi-local mass is shown for any $k$ without assumptions above.
We also classify all the vacuum solutions by utilizing the generalized Kodama vector.
Lastly, several conjectures on further generalization of the quasi-local mass in Lovelock gravity are proposed.
\end{abstract}



\maketitle


\section{Introduction}
In a general theory admitting a diffeomorphism invariance, the concept of local energy density becomes meaningless.
Expressions for mass-energy-momentum pseudo-tensors explicitly depending only on the metric and its first derivatives will vanish at any point of the spacetime in the locally flat coordinates~\cite{gravitation}.
This difficulty comes about as a natural result of the strong equivalence principle.
Thus we face a formidable issue arising in any theory of gravity derived by a diffeomorphism invariant Lagrangian. 
In spite of a considerable number of attempts to formulate a meaningful local energy density (see e.g.~\cite{pseudo,szabados2004} and references therein), we have not yet obtained an acceptable resolution to this problem.  
Localizing and identifying the gravitational mass-energy-momentum remains puzzling.

There exist, however, at least two satisfactory notions of total mass-energy (simply mass, hereafter) describing an isolated system in general relativity in four dimensions, that is the Arnowitt-Deser-Misner (ADM) mass~\cite{gravitation, ADM} and the Bondi mass~\cite{Bondi}.
Accordingly, it is tempting to employ the quasi-local mass~\cite{szabados2004,ms1964,by1993}, which is defined quasi-locally on the boundary of a given spacetime. 
For a finite region it contains a boundary term, which determines the boundary conditions and the value of a quasi-local mass.

From past studies of the quasi-local mass, it is suggested that a well-defined quasi-local mass should posses the five properties shown below~\cite{eardley1979}. 
(See~\cite{szabados2004} for a review.)
(i) When a two-sphere shrinks toward a point, the point in a spacetime must have zero mass.
(ii) A metric two-sphere in Minkowski spacetime should have zero mass.
(iii) In asymptotically flat spacetimes, it gives the ADM mass and the Bondi mass at spacelike and null infinities, respectively.
(iv) In spherically symmetric spacetimes, there exists a {\it mass function}, to which any definition of mass should reduce in the spherically symmetric case.
In particular, in Schwarzschild spacetime with the ADM or Bondi mass $M$, the mass function should give $M$.
(v) If a two-sphere $S$ is completely contained in the interior of another two-sphere $S'$, then the mass on $S'$ should be equal to or greater than the mass on $S$.

In the spherically symmetric case, the Misner-Sharp mass is widely
accepted as a well-posed quasi-local mass in general
relativity~\cite{ms1964}.\footnote{Here it should be noted that we have
another candidate for the quasi-local mass called the Brown-York
mass~\cite{by1993}. It is intimately related to the Hamilton-Jacobi
method and directly derived by the gravitational Hamiltonian. The
Brown-York mass satisfies the conditions (i)--(iii); however, it does
not reproduce the Misner-Sharp mass in the spherically symmetric
case. Thus, the uniqueness of the mass function in condition (iv) is
still an open problem up to now. Refer to \cite{blau} for a recent study.}
It satisfies the above conditions except for condition (v).
However, the condition (v) can be weakened to be satisfied only in the untrapped regions, and then the Misner-Sharp mass satisfies all the above conditions.
In spherically symmetric spacetimes, a very useful formulation of the basic equations in terms of the Misner-Sharp mass is available, with which it was shown that the Misner-Sharp mass is intimately related to the dynamical aspects of black holes and singularities~\cite{hayward1996}.
A generalization of the Misner-Sharp mass in the presence of a cosmological constant $\Lambda$ has also been considered, which inherits characteristics from the Misner-Sharp mass although the asymptotic structure of the spacetime is different from the case without $\Lambda$~\cite{nakao1995}.

In recent years, it has been of great importance to analyze physics in higher-dimensional spacetimes.
String theory is the most promising theory for unifying fundamental forces in nature and reduces to the higher-dimensional general relativity at the tree level. 
Even at the classical level, higher-dimensional gravity shows quite different aspects from that in four dimensions.
Studies of arbitrary dimensional gravity will reveal the characteristics of four-dimensional gravity.

In arbitrary dimensions, the most general action constructed from the Riemann curvature tensor and its contractions giving rise to the second-order quasi-linear field equations is given by the Lovelock polynomial~\cite{lovelock}. 
In four dimensions, it reduces to the Einstein-Hilbert action with $\Lambda$. 
Einstein-Gauss-Bonnet gravity, whose Lagrangian includes up to the quadratic term, arises in the low-energy limit of heterotic string theory as the higher curvature correction to general relativity~\cite{Gross}. 

The generalized Misner-Sharp mass in Einstein-Gauss-Bonnet gravity was recently proposed by one of the present authors~\cite{maeda2006b}.
In the vacuum case without $\Lambda$, it reduces to the higher-dimensional ADM mass in the unique spherically symmetric solution obtained by Boulware and Deser, and independently by Wheeler~\cite{GB_BH,Wheeler_1}.
Recently, it was shown that more pathological massive naked singularities, which are ruled out in general relativity, can be formed in five dimensions from the gravitational collapse of a physically reasonable matter in Einstein-Gauss-Bonnet gravity~\cite{maeda2006,maeda2006b}.
In their studies, the generalized Misner-Sharp mass was adopted to evaluate the mass of the singularities; however, the validity of that quasi-local mass has not been addressed so much.
The main purpose of the present paper is to fill this gap.

In this paper, we show that the generalized Misner-Sharp mass defined in~\cite{maeda2006b} is a natural counterpart of the Misner-Sharp mass in general relativity.
Our quasi-local mass agrees with a quasi-local conserved charge associated with a locally conserved current constructed from the generalized Kodama vector.
Using the simple mass variation formulae of the basic equations, we show that our quasi-local mass inherits characteristics from the Misner-Sharp mass such as monotonicity or positivity. 

The outline of the present paper is as follows.
Basic equations are given in the next section. 
In section \ref{sec:Kodamavector}, we discuss the relation between the generalized Kodama vector and our quasi-local mass. 
Section~\ref{sec:qlm} is devoted to investigating the properties of the quasi-local mass.
Our conclusions and discussions are summarized in section~\ref{sec:summary}, in which we propose a further generalization of the quasi-local mass in general Lovelock gravity and some associated conjectures.
Expressions of curvature tensors are given in appendix.
Conventions of curvature tensors are ${R^\mu }_{\nu\rho\sigma}V^\nu:=[\nabla _\rho ,\nabla_\sigma]V^\mu$ and $R_{\mu \nu }:={R^\rho }_{\mu \rho \nu }$.
The Minkowski metric is taken to be the mostly plus sign, and Roman indices run over all spacetime indices. 
We adopt units in which only the $n$-dimensional gravitational constant $G_n$ retained.

\section{Basic equations}
We begin by a brief description of Einstein-Gauss-Bonnet gravity 
in the presence of a cosmological constant.
The action in the $n (\geq 5)$-dimensional spacetime is given by
\begin{align}
\label{action}
S=\int\D ^nx\sqrt{-g}\biggl[\frac{1}{2\kappa_n^2}
(R-2\Lambda+\alpha{L}_{\rm GB}) \biggr]+S_{\rm matter},
\end{align}
where
$\kappa_n := \sqrt{8\pi G_n}$ and $R$ and $\Lambda$ are the $n$-dimensional Ricci scalar 
and the cosmological constant, respectively. 
$S_{\rm matter}$ in Eq.~(\ref{action}) is the action for matter fields.
The Gauss-Bonnet term $L_{\rm GB}$ comprises the combination of the Ricci scalar, 
Ricci tensor $R_{\mu\nu}$ and Riemann tensor ${R^\mu}_{\nu\rho\sigma}$
as
\begin{align}
{L}_{\rm GB} := R^2-4R_{\mu\nu}R^{\mu\nu}
+R_{\mu\nu\rho\sigma}R^{\mu\nu\rho\sigma}.
\end{align}
In the four-dimensional spacetime, the Gauss-Bonnet term 
does not contribute to the field equations since it
becomes a total derivative.
$\alpha$ is the coupling constant of the Gauss-Bonnet term. 
This type of action is derived in the low-energy limit 
of heterotic string theory~\cite{Gross}.
In that case, $\alpha$ is regarded as the inverse string tension 
and positive-definite. 
Thus, we also assume $\alpha \ge 0$ throughout this paper.
The gravitational equation of the action (\ref{action}) is
\begin{align}
{G^\mu}_{\nu} +\alpha {H}^\mu_{~~\nu} 
+\Lambda \delta^\mu_{~~\nu}= 
\kappa_n^2 {T}^\mu_{~~\nu}, \label{beq}
\end{align}
where 
\begin{align}
{G}_{\mu\nu}&:= R_{\mu\nu}-{1\over 2}g_{\mu\nu}R,\\
{H}_{\mu\nu}&:= 2\Bigl[RR_{\mu\nu}-2R_{\mu\alpha}
R^\alpha_{~\nu}-2R^{\alpha\beta}R_{\mu\alpha\nu\beta} \nonumber \\
&~~~~~~+R_{\mu}^{~\alpha\beta\gamma}R_{\nu\alpha\beta\gamma}\Bigr]
-{1\over 2}g_{\mu\nu}{L}_{\rm GB}
\end{align}
and ${T}^\mu_{~~\nu}$ is the energy-momentum tensor 
for matter fields obtained from $S_{\rm matter}$.
The field equations (\ref{beq}) contain up to the second derivatives 
of the metric.

Suppose the $n$-dimensional spacetime 
$({\ma M}^n, g_{\mu \nu })$ to be a warped product of an 
$(n-2)$-dimensional constant curvature space $(K^{n-2}, \gamma _{ij})$
and a two-dimensional orbit spacetime $(M^2, g_{ab})$ under 
the isometries of $(K^{n-2}, \gamma _{ij})$. Namely, the line element
is given by
\begin{align}
g_{\mu \nu }\D x^\mu \D x^\nu =g_{ab}(y)\D y^a\D y^b +r^2(y) \gamma _{ij}(z)
\D z^i\D z^j ,
\label{eq:ansatz}
\end{align} 
where
$a,b = 0, 1;~i,j = 2, ..., n-1$. 
Here $r$ is a scalar on $(M^2, g_{ab})$  with $r=0$ 
defining its boundary and $\gamma_{ij}$ is the unit
metric on $(K^{n-2}, \gamma _{ij})$ with its sectional curvature $k = \pm 1, 0$. 
We assume that $({\ma M}^n, g_{\mu \nu})$ is strongly causal and 
$(K^{n-2}, \gamma _{ij})$ is compact. Curvature tensors in this
spacetime are given in appendix~\ref{sec:curvature}.

The generalized Misner-Sharp mass in Einstein-Gauss-Bonnet gravity is a scalar function on $(M^2, g_{ab})$ 
with the dimension of mass such that
\begin{align}
\label{qlm}
m &:= \frac{(n-2)V_{n-2}^k}{2\kappa_n^2}
\biggl\{-{\tilde \Lambda}r^{n-1}
+r^{n-3}[k-(D r)^2] \nonumber \\
&~~~~~~+{\tilde \alpha}r^{n-5}[k-(Dr)^2]^2 \biggl\},
\end{align}  
where ${\tilde \alpha} := (n-3)(n-4)\alpha$, 
${\tilde \Lambda} := 2\Lambda /[(n-1)(n-2)]$,
$D_a $ is a metric compatible linear connection on $(M^2, g_{ab})$
and $(Dr)^2:=g^{ab}(D_ar)(D_br)$~\cite{maeda2006b}.
$V_{n-2}^k$ denotes the area of $K^{n-2}$.
In the four-dimensional spherically symmetric case 
without a cosmological constant, 
$m$ reduces to the Misner-Sharp quasi-local mass~\cite{ms1964}.

The line element may be written locally in the double-null coordinates as
\begin{align}
\D s^2 = -2e^{-f(u,v)}\D u\D v
+r^2(u,v) \gamma_{ij}\D z^i\D z^j. \label{coords}
\end{align}  
Null vectors $(\partial /\partial u)$ and $(\partial /\partial v)$ 
are taken to be future-pointing. 
The expansions of two independent future-directed radial null geodesics are defined as
\begin{align}
\theta_{+}&:=(n-2)r^{-1}r_{,v},\\
\theta_{-}&:=(n-2)r^{-1}r_{,u}.
\end{align}  
Here we give some definitions for later investigations.
\begin{dn}
\label{def:t-surface}
A {\it trapped (untrapped) surface} is an $(n-2)$-surface with $\theta_{+}\theta_{-}>(<)0$.
\end{dn}
\begin{dn}
\label{def:t-region}
A {\it trapped (untrapped) region} is the union of all trapped (untrapped) surfaces.
\end{dn}
\begin{dn}
\label{def:m-surface}
A {\it marginal surface} is an $(n-2)$-surface with $\theta_{+}\theta_{-}=0$.
\end{dn}
Observe that the value of $\theta _{+}$ or $\theta _{-}$ is not a geometrical
invariant because the null coordinates $u$ and $v$ have a rescaling
freedom $u\to U(u), v\to V(v)$. An invariant combination is 
$e^f\theta _+\theta _-$.
The function $r$, on the other hand, has a geometrical meaning
as an areal radius: the area of the symmetric subspace
is given by 
\begin{align}
A:=V^k_{n-2}r^{n-2}.
\label{area}
\end{align}
Then, the quasi-local mass $m$ is expressed as
\begin{align}
\label{qlm2}
m &= \frac{(n-2)V_{n-2}^k}{2\kappa_n^2}r^{n-3}
\biggl[-{\tilde \Lambda}r^2+\left(k+\frac{2}{(n-2)^2} r^2e^{f}
\theta_{+}\theta_{-}\right)\nonumber \\
&~~~~~~
+{\tilde \alpha}r^{-2}\left(k+\frac{2}{(n-2)^2} 
r^2e^{f}\theta_{+}\theta_{-}\right)^2\biggl].
\end{align}

The most general material stress-energy tensor $T_{\mu\nu}$ 
is given by
\begin{align}
T_{\mu\nu}\D x^\mu \D x^\nu =
&T_{uu}(u,v)\D u^2+2T_{uv}(u,v)\D u\D v \nonumber \\
&
+T_{vv}(u,v)\D v^2+p(u,v)r^2 \gamma_{ij}\D z^i\D z^j.
\end{align}  
By making use of the expressions given in appendix \ref{sec:curvature}, 
the governing equations~(\ref{beq}) are given by
\begin{widetext}
\begin{align}
&(r_{,uu}+f_{,u}r_{,u})\left[1+\frac{2{\tilde\alpha}}{r^2}
(k+2e^{f}r_{,u}r_{,v})\right]
=-\frac{\kappa_n^2}{n-2} r T_{uu}, \\
&(r_{,vv}+f_{,v}r_{,v})\left[1+\frac{2{\tilde\alpha}}{r^2}
(k+2e^{f}r_{,u}r_{,v})\right]
=-\frac{\kappa_n^2}{n-2} r T_{vv}, \label{equation:vv} \\
&rr_{,uv}+(n-3)r_{,u}r_{,v}+\frac{n-3}{2}k e^{-f}+\frac{{\tilde\alpha}}{2r^2}
[(n-5)k^2e^{-f}+4rr_{,uv}
(k+2e^fr_{,u}r_{,v})+4(n-5)r_{,u}r_{,v}
(k+e^fr_{,u}r_{,v})] \nonumber \\
&~~~~~~-\frac{n-1}{2}{\tilde\Lambda}r^2e^{-f}
=\frac{\kappa_n^2}{n-2} r^2T_{uv},\\
&r^2 f_{,uv}+2(n-3)r_{,u}r_{,v}
+k(n-3)e^{-f}-(n-4)rr_{,uv} \nonumber \\
&~~~~~~+\frac{2{\tilde\alpha}e^{-f}}{r^2}
\biggl[e^f(k+2e^fr_{,u}r_{,v})
\{r^2f_{,uv}-(n-8)rr_{,uv}\}
+2r^2e^{2f}(f_{,u}r_{,u} r_{,vv}
+f_{,v}r_{,v}r_{,uu}) \nonumber \\
&~~~~~~+(n-5)(k+2e^fr_{,u}r_{,v})^2
+2r^2e^{2f}\{r_{,uu}r_{,vv}
+f_{,u}f_{,v}r_{,u}r_{,v}
-(r_{,uv})^2\}\biggl] \nonumber \\
&~~~~~~=\kappa_n^2 r^2(T_{uv}+e^{-f}p).
\end{align}  
\end{widetext}

The variation of $m$ is determined by these equations as
\begin{align}
m_{,v}&=
\frac{1}{n-2}V_{n-2}^ke^fr^{n-1}(T_{uv}\theta_+-T_{vv}\theta_-), \label{m_v} \\
m_{,u}&=
\frac{1}{n-2}V_{n-2}^ke^fr^{n-1}(T_{uv}\theta_- -T_{uu}\theta_+). \label{m_u} 
\end{align}  
These variation formulae are exactly the same 
as those in general relativity, which enable us to 
prove most of the lemmas and propositions in this paper in close parallel with the
general relativistic case.

Instead of specifying the matter fields, energy conditions
are imposed in the present paper.
The null energy condition for the matter field implies
\begin{align}
T_{uu}\ge 0,~~~T_{vv} \ge 0, \label{nec}
\end{align}
while the dominant energy condition implies
\begin{align}
T_{uu} \ge 0,~~T_{vv}\ge 0,~~T_{uv}\ge 0, \label{dec}
\end{align}  
which assures that a causal observer measures the 
non-negative energy density and the energy flux is 
a future-directed causal vector.

\section{Generalized Kodama vector and quasi-local mass}
\label{sec:Kodamavector}

In this section, we explicitly show that $m$ is a quasi-local 
conserved quantity associated with a locally conserved current.
First we give the definition of the generalized Kodama vector~\cite{kodama1980,ms2004}:
\begin{align}
K^\mu  :=-\epsilon ^{\mu \nu }\nabla _\nu  r,
\label{kodamavector}
\end{align}
where 
$\epsilon_{\mu \nu}=\epsilon_{ab}(\D x^a)_{\mu}(\D x^b)_{\nu}$,
and $\epsilon_{ab}$
is a volume element of $(M^2, g_{ab})$. 
In the double null coordinates (\ref{coords}), we have $\epsilon_{uv}=e^{-f}$ and $\epsilon^{uv}=-e^{f}$.
The Kodama vector was originally introduced in four-dimensional 
spherically symmetric spacetimes~\cite{kodama1980}.

In the double null coordinates, we have
\begin{eqnarray}
\label{kodama}
K^\mu\frac{\partial}{\partial x^\mu} = e^f 
\biggl(r_{,v}\frac{\partial}{\partial u}- r_{,u}\frac{\partial}{\partial v}\biggl).
\end{eqnarray}  
It follows immediately that 
$K^\mu $ is tangent to $\{r={\rm const.}\}$ surfaces, i.e., 
$K^\mu $ and $\nabla^\mu r$ are orthogonal
\begin{align}
K^\mu \nabla _\mu r=0.
\end{align}
This feature illustrates that 
$K^\mu $ is the analogue of the Hamiltonian vector field
with an energy function $r$ on a symplectic manifold.
It is also shown that 
\begin{eqnarray}
K^\mu  K_\mu =-(\nabla r)^2=2e^fr_{,v}r_{,u},
\end{eqnarray}  
so that $K^\mu$ is timelike and spacelike in the untrapped and 
trapped region, respectively, and it is null on marginal surfaces.
In the untrapped region, $K^\mu $ generates a preferred time evolution.
The minus sign in the right side of (\ref{kodamavector}) ensures
that $K^\mu $ is {\it future-directed} in the untrapped region.

Since two orthogonal null vectors are proportional to each other, 
we have
\begin{align}
K^\mu =\nabla ^\mu r,
\label{proportional}
\end{align} 
on marginal surfaces,
where the proportionality factor has been determined 
so as to be consistent with ({\ref{kodamavector}}).

By definition, we readily see that $K^\mu $ is a local conserved current 
\begin{align}
\nabla _\mu K^\mu =0.\label{divK}
\end{align}
It is also shown by direct calculations that  
\begin{align}
G_{\mu\nu}\nabla ^\mu K^\nu&=0,\label{divG} \\
H_{\mu\nu}\nabla ^\mu K^\nu &=0 \label{divH}
\end{align}  
hold, where we have used expressions in appendix~\ref{sec:curvature} together with
$\nabla_\mu K_\nu=D_aK_b(\D x^a)_{\mu}(\D x^b)_{\nu}$ and $(D_aD_b r)D^aK^a=0$.

Equations~(\ref{divK}), (\ref{divG}) and (\ref{divH}) imply that the vector fields
\begin{align}
J_{(0)}^\mu &:= -\frac12g^{\mu\nu}K_{\nu},\\
J_{(1)}^\mu &:= G^{\mu\nu}K_{\nu},\\
J_{(2)}^\mu &:= H^{\mu\nu}K_{\nu}
\end{align}  
are also divergence-free because of the identities 
$\nabla _\nu g^{\mu\nu}\equiv 0$, 
$\nabla _\nu G^{\mu\nu}\equiv 0$ 
and $\nabla _\nu H^{\mu\nu}\equiv 0$. 
Thus, three independent locally conserved currents  
$J_{(0)}^\mu$, $J_{(1)}^\mu$ and $J_{(2)}^\mu$ 
are constructed from the generalized Kodama vector $K^\mu$. 
Here we define
\begin{align}
J^\mu := -\frac{1}{\kappa _n^2}
\left( -2\Lambda J_{(0)}^\mu +J_{(1)}^\mu +\alpha J_{(2)}^\mu  \right),
\label{kodamacurrent}
\end{align}
which is also divergence-free
\begin{align}
 \nabla_\mu J^\mu=0.
\label{divJ}
\end{align}
Each coefficient in Eq. (\ref{kodamacurrent}) was chosen such that,
 by virtue of field equations, $J^\mu=-{T^\mu}_\nu K^\nu$
representing the energy current.

Since $J^\mu$ is divergence-free (\ref{divJ}),
there exists, at least locally, a potential function $\Phi$ such that
\begin{align}
J^\mu =-\epsilon ^{\mu \nu}\nabla_\nu \Phi.
\end{align} 
Namely, $J^\mu$ is a Hamiltonian vector field with an
energy function $\Phi$. 
In the untrapped region, $J^\mu $ is a future-directed 
causal vector if the dominant energy
condition holds. 
The integrals of locally conserved currents $K^\mu $ and $J^\mu $ 
over some spatial volume $\Sigma $ with boundary
give associated charges: 
\begin{align}
Q_K &:=\int _\Sigma K^\mu \D \Sigma _\mu , \\
Q_J &:= \int _\Sigma J^\mu \D \Sigma _\mu,
\label{kodamamass}
\end{align}
where $\D \Sigma _\mu$ is a directed surface element on $\Sigma $.
If $\Sigma $ has no boundary, 
these quantities will be independent of the choice of $\Sigma $ when
$\Sigma $ is compact or the integrand vanishes at infinity.

Now we introduce the coordinates as
\begin{align}
\D s^2 =-e^{2\phi (t,\rho)}\D t^2+e^{2\psi (t,\rho)}\D \rho^2
+r^2(t,\rho)\gamma _{ij}\D z^i\D z^j
\label{t-xcoordinate}
\end{align}
and take the spatial volume $\Sigma $ as 
$\Sigma =\{t=t_0={\rm const.}, 0\le \rho\le \rho'\}$. 
In this set of coordinates, we have $\epsilon_{t\rho}=e^{\phi+\psi}$ and $\epsilon^{t\rho}=-e^{-\phi-\psi}$, so that  
\begin{eqnarray}
K^\mu\frac{\partial}{\partial x^\mu} = e^{-\phi-\psi}\biggl(r_{,\rho}\frac{\partial}{\partial t}- r_{,t}\frac{\partial}{\partial \rho}\biggl).
\end{eqnarray}  
A future-directed unit normal to $\Sigma $ is then $u^\mu :=e^{-\phi}(\partial /\partial t)^\mu$ and a directed surface element is written by $u^\mu$ and a surface element $\D \Sigma$ as $\D \Sigma _\mu=-u_\mu \D \Sigma$.
Then, it is
a tedious but straightforward task to show
\begin{align}
Q_K&=V_{n-2}^kr^{n-1}/(n-1),
\label{volume2} \\
Q_J&=m,
\label{QJ=m}
\end{align}
where $r$ and $m$ are evaluated at $t=t_0$ and $\rho=\rho'$.
Because the areal volume $V$ is defined by
\begin{align}
V:=V_{n-2}^kr^{n-1}/(n-1),\label{volume}
\end{align}
$Q_K$ is interpreted as the areal volume and
actually divergent for a non-compact $\Sigma $.
The values of $Q_K$ and $Q_J$, of course, depend on the
particular choice of $\Sigma $, reflecting their quasi-local
nature.
Eq. (\ref{QJ=m}) is the main result in this section.
It should be observed that although arbitrary linear combinations
of $J^\mu_{(0)}, J^{\mu}_{(1)}$ and $J^\mu_{(2)}$ give locally conserved
currents, only the energy current form (\ref{kodamacurrent}) is
associated with our quasi-local mass.

The existence of a symmetry entails the conserved Noether charge
as a symmetry generator.
More precisely, Noether's theorem states that the invariance of the 
Hamiltonian $H$ along a vector field $\xi^\mu$ implies the conserved
charge $Q_\xi$ 
through the Poisson bracket
\begin{align}
0=\{H, Q_\xi\}_{\rm PB}=\mathscr L_\xi H.
\label{noether}
\end{align}
Now $\epsilon _{\mu \nu }$ is a closed two-form, it is identified as 
a symplectic structure.
Let us see the above in the language of symplectic structure 
(see e.g.,  \cite{arnold}) and
further discuss the relation between conserved currents and
associated charges. 
The symplectic structure $\epsilon_{\mu \nu}$
naturally induces the Poisson bracket
\begin{align}
\{A, B\}_{\rm PB}&:=-\epsilon^{\mu \nu} (\nabla_\mu A)(\nabla_\nu B), \\
&=V_B^\mu \nabla_\mu A =\mathscr L_{V_B}A,
\end{align}
where $A$ and $B$ are scalar functions on $(M^2, g_{ab})$, and
$V_B^\mu:=-\epsilon^{\mu \nu}(\nabla_\nu B)$ is a Hamiltonian vector field associated with $B$.
If we take $A$ as the Hamiltonian and $B$ as a charge associated with a
vector $\xi^\mu$, we reproduce Eq. (\ref{noether}).
Using the above formula, we calculate the Poisson bracket between charges
and associated energy functions. We obtain
\begin{align}
0=\{V, r\}_{\rm PB}=\mathscr L_K V=K^\mu\nabla_\mu V,
\end{align}
and
\begin{align}
0=\{m, \Phi\}_{\rm PB}=\mathscr L_J m=J^\mu \nabla_\mu m,
\end{align}
both of which show that $V$ and $m$ are conserved along $K^\mu$ and
$J^\mu$, respectively.

\section{Properties of the quasi-local mass}
\label{sec:qlm}
In this section, properties of the quasi-local mass (\ref{qlm}) 
such as the energy balance law, vacuum, asymptotic behavior, 
monotonicity and positivity, are examined.

\subsection{Unified first law}
The first law of thermodynamics is one of the elementary laws of physics
representing an energy conservation. 
Thus, the first law can be used as an explicit criterion concerning the properness of the definition of mass.  
We will show that this is indeed the case for the quasi-local mass as well: it satisfies the unified first law.

We define a scalar 
\begin{align}
P:=-\frac 12 {T^a}_a
\end{align}
and a vector
\begin{align}
\psi ^a :={T^a}_bD ^b r +PD ^a r
\end{align}
on $(M^2, g_{ab})$,
where the contraction is taken over on the two-dimensional orbit space.
The areal volume $V$ given by (\ref{volume}) satisfies $D_a V=AD _ar$, where $A$ is given by (\ref{area}).
By using the field equations (see 
equations in appendix \ref{sec:curvature}) and
utilizing the identity (\ref{S=0}),
we obtain
\begin{align}
\D  m =A\psi_a\D x^a  +P\D V.
\label{1stlaw1}
\end{align}
This is the unified first law
corresponding to an energy balance law~\cite{hayward1998}.
The first term in the right-hand-side represents an energy flux, 
while the second does an external work~\cite{hayward1998,ashworth1999}. 
Assuming the dominant energy condition, 
we have $P\ge 0$.
In the double null coordinates, the unified first law
gives the variation formulae (\ref{m_v}) and (\ref{m_u}).

\subsection{Vacuum}
In the vacuum case, it follows from Eq. (\ref{m_v})
and (\ref{m_u}) that $m_{,u}=m_{,v}=0$,
i.e., $m=M$, where $M$ is a constant. 
A static vacuum solution, which we call the generalized Boulware-Deser-Wheeler solution~\cite{GB_BH,Wheeler_1,EGBBH,tm2005}, is given by
\begin{align}
\D s^2=-F(r)\D t^2+F^{-1}(r)\D r^2+r^2\gamma_{ij}\D z^i
\D z^j,\label{BDW1}
\end{align}
where
\begin{align}
\label{BDW}
F(r) := k+\frac{r^2}{2\tilde{\alpha }}\left[
1\mp \sqrt{1+\frac{8\kappa _n^2\tilde{\alpha }M}
{(n-2)V_{n-2}^kr^{n-1}}+4{\tilde\alpha}{\tilde\Lambda}}\right].
\end{align}

In the case where $k=1$ and $\Lambda=0$, the staticity assumption is redundant and the generalized Birkhoff's theorem holds, namely the Boulware-Deser-Wheeler solution~(\ref{BDW1}) is the general solution~\cite{Wiltshire:1985us}.
For general $k$ and $\Lambda$, on the other hand, other solutions are possible.
We classify all the vacuum solutions below by utilizing the generalized Kodama vector.
The following proposition includes the results in~\cite{Wiltshire:1985us,cd2002,dt2003} and a special case of the results in~\cite{zegers2005,df2005} in Lovelock gravity, in which $(Dr )^2 \ne 0$ is implicitly assumed.

\begin{Prop}
\label{th:vacuum}
({\it Vacuum solutions.})
An $n$-dimensional vacuum spacetime in Einstein-Gauss-Bonnet
gravity with the metric form (\ref{eq:ansatz}) is isometric to one of the followings:
(i) the generalized Boulware-Deser-Wheeler solution (\ref{BDW1}) if $(Dr )^2 \ne 0$,
(ii) the Nariai-type solution (\ref{Nariai}) if $r$ is constant, and
(iii) the solution (\ref{eq:staticmetric}) if $(Dr)^2=k+r^2/(2\tilde \alpha)$.
\end{Prop}
\noindent
{\it Proof.}
For the warped product spacetime (\ref{eq:ansatz}), independent 
vacuum field equations are given by
\begin{widetext}
\begin{align}
\left[1+\frac{2\tilde \alpha }{r^2}[k-(Dr)^2]\right]{}^{(2)}
R-(n-2)\frac{D^2 r}{r}-
2(n-1)\tilde \Lambda 
+\frac{2\tilde{\alpha }}{r^2}\biggl[
&[k-(Dr)^2]\biggl\{-\frac{(n-6)D^2 r}{r}-
\frac{(n-5)[k-(Dr)^2]}{r^2} \biggr\} \nonumber \\
& +2[(D^2 r)^2-(D_aD_br)(D^aD^b r)]
\biggr]=0,
\label{eq:1}
\end{align}
%
\begin{align}
\left(1+\frac{2 \tilde{\alpha }}{r^2}[k-(Dr)^2]\right)
\biggl(D_a D_b r-\frac 12 g_{ab}D^2 r \biggr)
=0,
\label{eq:2}
\end{align}
%
\begin{align}
-\frac{D^2 r}{r}+(n-3)\frac{k-(Dr)^2}{r^2}-
(n-1)\tilde \Lambda 
+\frac{2 \tilde{\alpha }[k-(Dr)^2]}{r^2}\biggl[
\frac{(n-5)[k-(Dr)^2]}{2 r^2}-\frac{D^2 r}r\biggr]=0.
\label{eq:3}
\end{align}
\end{widetext}
In deriving Eq. (\ref{eq:2}), 
we have used the two-dimensional identity (\ref{S=0}).
%
%
%
Eq. (\ref{eq:2}) requires either
\begin{align}
\textrm{class I:} \qquad 
1+\frac{2 \tilde{\alpha }}{r^2}[k-(Dr)^2]=0
\label{classI}
\end{align}
or
\begin{align}
\textrm{class II:} \qquad 
D_a D_b r-\frac 12 g_{ab}D^2 r=0.
\label{classII}
\end{align}

We first analyze the class I.
Substituting (\ref{classI}) into Eq. 
(\ref{eq:3}) yields
\begin{align}
1+4\tilde \alpha \tilde \Lambda =0.
\label{DCgravity}
\end{align}
Together with (\ref{classI}), this implies the vanishing of quasi-local mass $m \equiv 0$. 
From Eqs. (\ref{eq:1}) and (\ref{DCgravity}), we have
\begin{align}
\frac{D^2 r}r =\frac 1{2\tilde \alpha }+\frac{\tilde \alpha }{r^2}
[(D^2 r)^2-(D_aD_b r)(D^a D^b r)].
\end{align}
If $r=r_0={\rm const.}$, or if $D^a r$ is null, 
it leads to a contradiction.
If $(Dr)^2\ne 0$, we find a general solution by choosing 
$r$ as one of the coordinates: 
\begin{align}
\D s^2=-h(r)e^{2\delta (t, r)}\D t^2 +{h^{-1}(r)}
\D r^2+r^2 \gamma _{ij}\D z^i\D z^j,
\label{eq:staticmetric}
\end{align}
where $h(r):=k+r^2/(2\tilde \alpha )$, $1+4\tilde \alpha \tilde \Lambda =0$ and $\delta (t, r)$ is an {\it arbitrary } function. 
Hence the class I solution is not static in general.
If $\delta =\delta (t)$, this corresponds to 
the dimensionally extended constant curvature black hole given by Ban\~ados, Teitelboim and Zanelli \cite{BTZ}.

We next analyze class II. We first note that
Eq. (\ref{classII}) implies that $D^a r$ is a 
conformal Killing vector on $(M^2, g_{ab})$. 
We find from Eqs. (\ref{classII}) and the definition of 
the generalized Kodama vector (\ref{kodama}) that 
\begin{align}
D_a K_b = -\frac 12 \epsilon _{ab}D^2 r, \label{dK}
\end{align}
which in turn implies that $K^a $ is a Killing vector field 
on $(M^2, g_{ab})$, i.e., $D_{(a}K_{b)}=0$. Since 
$\nabla _\mu K_\nu =D_aK_b(\D x^a)_{\mu}(\D x^b)_{\nu}$,
{\it we conclude that $K^\mu $ is a hypersurface-orthogonal 
Killing vector on $({\ma M}^n, g_{\mu \nu })$}:
\begin{align}
K_{[\mu }\nabla _\nu K _{\rho ]}=0, \qquad
\nabla _{(\mu }K _{\nu )}=0.
\label{statickilling}
\end{align}

If $D^a r$ is a null vector, we can choose $r=u$ or $r=v$ without loss of generality.
Then, from Eqs.~(\ref{eq:1}) and (\ref{eq:3}), only the case of $k=0$ with $\Lambda=0$ is allowed, and consequently $m \equiv 0$ is given from Eq.~(\ref{qlm}).
For $r=u$, Eq.~(\ref{classII}) gives 
\begin{align}
\D s^2=-2\D u \D v+u^2\delta_{ij} \D z^i \D z^j,
\label{wave}
\end{align}
which is the Minkowski solution written in null coordinates. 
For $r=v$, we obtain the solution with $u$ and $v$ interchanged:
again reproduces the flat space.

Next we consider the case in which $D^a r$ is not null. 
Suppose first the generalized Kodama vector is timelike. Due to its 
hypersurface-orthogonality (\ref{statickilling}), we can choose 
$K^\mu =(\partial /\partial t)^\mu $ in the coordinates
(\ref{t-xcoordinate}), and then all the metric components 
($\phi, \psi$ and $r$) are independent of $t$. 
The unified first law (\ref{1stlaw1}) implies that $m$
is constant, and
Eq. (\ref{classII}) now reduces to
\begin{align}
\frac{\D}{\D \rho}(\phi+\psi )=0, \qquad 
\frac{\D^2}{\D \rho^2}r=0,
\end{align} 
or 
\begin{align}
r=r_0={\rm constant}.
\end{align}

In the former case, the remaining gauge degrees of freedom enable us to set
$\phi=-\psi$ and $r=\rho$. 
Finally, Eq.~(\ref{eq:1}) or (\ref{eq:3}) indicates that the resulting spacetime is isometric to
the generalized Boulware-Deser-Wheeler solution (\ref{BDW1}).  

In the latter case of $r=r_0={\rm const. }$, Eq.~(\ref{eq:1}) gives that ${}^{(2)}R$ is constant, i.e., $M^2$ is a two-dimensional constant curvature spacetime, which is the two-dimensional flat, de~Sitter, or anti-de~Sitter spacetime.
Thus, ${\ma M}^n$ is the Nariai-type spacetime, of which the metric is given in the standard coordinates as~\cite{md2007}:
\begin{align}
 \D s^2 =-(1-\sigma \rho^2)\D t^2+\frac{\D \rho^2}
{1-\sigma \rho^2}+r_0^2\gamma _{ij}\D z^i\D z^j,
\label{Nariai}
\end{align}
where
\begin{align}
\sigma :=\left[\frac{2(n-3)+2\tilde \alpha 
(n-5)kr_0^{-2}}{r_0^2+2\tilde \alpha k}\right]k
\end{align}
and $r_0^2$ is the real and positive root of the following algebraic equation: 
\begin{align}
(n-1)\tilde{\Lambda }=\frac{(n-3)k}{r_0^2}+
\frac{(n-5)\tilde \alpha k^2}{r_0^4}.
\end{align}
The existence condition of the real and positive $r_0^2$ is $\Lambda>0$ and $k=\pm1$ or $-(n-3)^2/[4(n-1)(n-5){\tilde\alpha}]\le {\tilde\Lambda} \le 0$ and $k=-1$ for $n \ge 6$, while it is $k{\tilde\Lambda}>0$ for $n=5$.

If the generalized Kodama vector is spacelike, the Nariai-type solution (\ref{Nariai}) or the dual ``interior'' solution of (\ref{BDW1}), i.e., the solution with $t$ and $r$ interchanged, is obtained. 
\qed

\bigskip

It is noted that the condition $(Dr)^2=k+r^2/(2\tilde \alpha)$ in Proposition~\ref{th:vacuum} inevitably leads to a special relation between coupling constants (\ref{DCgravity}), but its inverse does not hold.
Actually, the generalized Boulware-Deser-Wheeler solution (\ref{BDW1}) with any $k$ and the Nariai-type solution (\ref{Nariai}) with $k=-1$ also admit the special relation (\ref{DCgravity}).
Five-dimensional Einstein-Gauss-Bonnet gravity with the relation (\ref{DCgravity}) is a class of Chern-Simons gravity defined in odd dimensions~\cite{Banados:1993ur,zanelli2005}.

\subsection{Asymptotic behavior}
\label{subsec:asymp}

We next discuss the asymptotic property of the quasi-local 
mass in asymptotically flat spacetimes.
It is shown that the quasi-local mass $m$ 
gives the ADM mass at spatial infinity.
 
\begin{Prop}
\label{th:asymptotics}
({\it Asymptotic behavior in asymptotically flat spacetime.}) 
In an $n$-dimensional asymptotically flat spacetime, 
$m$ coincides with the higher-dimensional ADM mass at spatial infinity.
\end{Prop}
\noindent
{\it Proof}.
In an $n$-dimensional  asymptotically flat spacetime, 
there exists a coordinate
system such that
\begin{align}
\D s^2 &\simeq -\left [1-\frac{2\kappa _n^2 M}{(n-2)
\ma A_{n-2}\rho ^{n-3}} \right]\D t^2 
-\frac{\kappa _n^2 J_{ij}x^i}{\ma A_{n-2}\rho ^{n-1}}
 \D x ^j \D t \nonumber \\
&
+ \left[1+\frac{2\kappa _n^2 M}{(n-2)(n-3)
\ma A_{n-2}\rho ^{n-3}}\right]\D x^i\D x^i,
\label{asy-coordinate}
\end{align}
around spatial infinity $\rho \to \infty $, 
where $\rho :=\sqrt{\sum_{i=1}^{n-1}(x^i)^2}$ is defined on
an $(n-1)$-dimensional Euclidean space~\cite{myersperry1986}. 
$\ma A_{n-2}$ is the surface area of an $(n-2)$-dimensional unit sphere
\begin{align}
\ma A_{n-2} :=\frac{2\pi^{(n-1)/2}}{\Gamma((n-1)/2)},
\label{unitarea}
\end{align}
where $\Gamma(x)$ is the Gamma function.
The constants $M$ and $J_{ij}$ are the higher-dimensional 
ADM mass and the higher-dimensional ADM angular momenta, 
respectively, where the number of components of $J_{ij}$ is given by the
integer part of $(n-1)/2$ corresponding to the rank of SO($n-1$).
Consequently, the areal coordinate $r$ asymptotically takes the value
\begin{align}
r\simeq \rho  \left[1+\frac{\kappa _n^2 M}
{(n-2)(n-3)\ma A_{n-2}\rho ^{n-3}}\right].
\end{align}
Substituting this into Eq. (\ref{qlm}) with $k=1$,
$V^k_{n-2}=\ma A_{n-2}$ and $\Lambda =0$, we have
\begin{align}
m|_{\rho \to \infty }=M.
\end{align}
\qed

\bigskip

The above proposition can be also shown from the result in the previous section.
Let the spatial volume $\Sigma $ extend out to the spacelike infinity.  
Since the spatial part of the generalized Kodama vector vanishes and it reduces 
to a timelike Killing vector asymptotically in the asymptotically flat
spacetime, the charge (\ref{kodamamass}) is strictly conserved independent of time-slicing. 
Thus Eq. (\ref{QJ=m}) provides the identical result because in the asymptotically flat case, higher-order curvature terms fall off sufficiently rapidly at infinity, so that they do not contribute to the conserved charges such as $M$ or $J_{ij}$. 
(See the expressions in \cite{iyerwald1994,antonio2003,deser2002,Deruelle:2003ps}.)

It deserves to be noted here on the asymptotic behavior of the
quasi-local mass (\ref{qlm}) at null infinity in the asymptotically flat spacetimes.
The Misner-Sharp mass is asymptotic to the Bondi mass at null infinity in general relativity~\cite{hayward1996}.
This asymptotic behavior is one of the criteria for the well-posedness of a quasi-local mass.
Thus, our quasi-local mass should be asymptotic to the higher-dimensional Bondi mass in that limit.
However, as demonstrated in~\cite{hollands2005,hollands2004}, 
we cannot define the Bondi-like radiation energy in an asymptotically 
flat spacetime in {\it odd} dimensions due to the absence of a stable
conformal null infinity.\footnote{This peculiar characteristic in odd dimensions 
may be related to the late-time behavior of the gravitational radiation~\cite{cardoso2003}.}
In the vacuum case, the mass parameter $m$ in the Boulware-Deser-Wheeler solution gives the higher-dimensional ADM mass at spacelike infinity and coincides with the higher-dimensional Bondi mass at null infinity in even dimensions as well, because the higher curvature terms fall off sufficiently rapidly also at null infinity .
But it is not clear whether the odd dimensional expression of the higher-dimensional Bondi mass is meaningful in its own right.

Next, we investigate the value of our quasi-local mass 
in the asymptotically anti-de Sitter (AdS) region.
We employ the asymptotically AdS boundary condition of Henneaux and Teitelboim for the metric
components adopting the coordinates $x^\mu=\{t, \rho , z^i\}$~\cite{HT1985}. (See~\cite{HIM2005} for the higher-dimensional version.) 
The metric under consideration can be written as
$g_{\mu \nu }=g^{(0)}_{\mu \nu}+h_{\mu \nu}$, where $g^{(0)}_{\mu \nu}$ is the metric of the AdS spacetime, from which deviation is represented by $h_{\mu \nu}$.
In the global coordinates, we have
\begin{align}
g^{(0)}_{\mu \nu } \D x^\mu \D x^\nu &=-\left(1+\frac{\rho ^2}{\ell_{\rm eff }^2}\right)\D t^2+
\left(1+\frac{\rho ^2}{\ell_{\rm eff }^2}\right)^{-1}\D \rho ^2 \nonumber \\
&~~~~~~~~~~~~~~~~~~+\rho ^2\D \Omega _{n-2}^2,
\label{AdS}
\end{align}
where $\D\Omega _{n-2}^2$ is 
the line element of a unit $(n-2)$-sphere.
The effective curvature radius in this spacetime is given by 
\begin{align}
\ell_{\rm eff}^2:=-\frac{1}{2\tilde \Lambda }\left(1\pm \sqrt{1+4\tilde \alpha
 \tilde \Lambda }\right).
\end{align}
The AdS spacetime (\ref{AdS}) solves the vacuum field equations corresponding to the generalized
Boulware-Deser-Wheeler solution (\ref{BDW1}) with $k=1$ and $M=0$.
The fall-off condition is such that 
\begin{subequations}
\begin{align}
&h_{tt}={c_1}\rho ^{-n+3}+O(\rho ^{-n+2}), \\ 
&h_{\rho \rho }={c_2}{\rho ^{-n-1}}+O(\rho ^{-n-2}), \\
&h_{t\rho }=c_3\rho ^{-n}+O(\rho ^{-n-1}), \\
& h_{\rho i}=c_4\rho ^{-n}+O(\rho ^{-n-1}),\\
&h_{ti}=c_5\rho ^{-n+3}+O(\rho ^{-n+2}), \\
&h_{ij}=c_6\rho ^{-n+3}+O(\rho ^{-n+2}),
\end{align}
\label{bc}
\end{subequations}
where $c_1,...,c_6$ are functions independent of $\rho $. 
In the $n$-dimensional spherically symmetric spacetime,
 which is of our interest here, $c_1,...,c_6$ are independent of $z^i$ and $c_4=c_5=0$.

Using the gravitational Hamiltonian formalism \cite{HH},
Padilla gave an expression of the global mass-energy in Einstein-Gauss-Bonnet gravity 
for the maximally symmetric background as \cite{Padilla2003}
\begin{align}
E:= -\frac{\Xi_{\rm q}}{\kappa _n^2}\int _{S}N(K-K_0)\D S, \label{p-mass}
\end{align}
where $\Xi_{\rm q} :=\pm \sqrt{1+4\tilde \alpha \tilde \Lambda }$ and $N$ is the lapse function.
We call $E$ the Padilla mass. 
Here $K$ is the extrinsic curvature of $(n-2)$-sphere $S$
at infinity with respect to a spatial surface $\Sigma$. 
 $K_0$ is the extrinsic curvature of  $(n-2)$-sphere 
with the {\it identical} intrinsic geometry embedded in the 
background AdS space (\ref{AdS}).  
The Padilla mass~(\ref{p-mass}) reproduces the Deser-Tekin mass, i.e., the global mass-energy obtained as a Killing charge~\cite{deser2002}.
(See also \cite{Deruelle:2003ps,OK2005} for comparison.)

We use the coordinates (\ref{t-xcoordinate}) and take the spatial surface such as
$\Sigma =\{t={\rm const.}\}$. Then
we have
$N\simeq \rho/\ell_{\rm eff }$ and 
\begin{align}
K&\simeq  \frac{(n-2)}{\ell_{\rm eff}}\left[1-\frac{(n-1)c_6}{2\rho^{n-1}}\right]
\left[1+\frac{\ell_{\rm eff}^2}{2\rho^2}-\frac{c_2\ell_{\rm eff}^{-2}}{2\rho^{n-1}}\right], \\
K_0&\simeq \frac{(n-2)}{\ell_{\rm eff}}\left[1-\frac{(n-1)c_6}{2\rho^{n-1}}\right]
\left[1+\frac{\ell_{\rm eff}^2}{2\rho^2}\right].
\end{align}
Putting all together, we arrive at
\begin{align}
E=\pm \frac{(n-2)A_{n-2}c_2}
{2\kappa_n^2\ell_{\rm eff}^{4}}\sqrt{1+4\tilde \alpha \tilde \Lambda },
\label{gravE}
\end{align}
where $\mathcal A_{n-2}$ is the area of unit $(n-2)$-sphere 
given by (\ref{unitarea}).

It is shown that our quasi-local mass $m$ approaches~(\ref{gravE}) at infinity.
\begin{Prop}
\label{th:asymptotics2}
({\it Asymptotic behavior in asymptotically AdS spacetime.}) 
In an $n$-dimensional asymptotically AdS spacetime, 
$m$ coincides with the Padilla and Deser-Tekin mass at infinity.
\end{Prop}
\noindent
{\it Proof}.
Substituting the asymptotic boundary conditions 
(\ref{bc}) into the definition of our quasi-local mass (\ref{qlm})
for $k=1$ and $V^k_{n-2}=\mathcal A_{n-2}$, 
we obtain
\begin{align}
m|_{\rho \to \infty}&=\frac{(n-2)
\mathcal A_{n-2}c_2}{2\kappa^2_n\ell^{4}_{\rm eff}}
\biggl(1-\frac{2{\tilde\alpha}}{\ell^{2}_{\rm eff}}\biggl),\nonumber \\
&=\pm\frac{(n-2)\mathcal A_{n-2}c_2}{2\kappa^2_n\ell^{4}_{\rm eff}}
\sqrt{1+4\tilde \alpha \tilde \Lambda },\nonumber \\
&=E,
\end{align}
where we used the fact $h^{\rho\rho}\simeq -c_2\ell^{-4}_{\rm eff}\rho^{-n+3}$ for $\rho \to \infty$.
\qed

\subsection{Monotonicity and positivity}
In this subsection, we investigate two important properties
 of the quasi-local mass $m$, namely monotonicity and positivity.
We fix the orientation of the untrapped surface by
$\theta _+>0$ and $\theta_-<0$, i.e., $\partial/\partial u$ and $\partial/\partial v$ are ingoing and outgoing null vectors, respectively.

\begin{Prop}
\label{th:monotonicity}
({\it Monotonicity.}) 
If the dominant energy condition holds, 
$m$ is non-decreasing (non-increasing) in any outgoing  
(ingoing) spacelike or null direction on an untrapped surface.
\end{Prop}
\noindent
{\it Proof}. 
Let $s^\mu(\partial/\partial x^\mu)=s^v(\partial/\partial v)+s^u(\partial/\partial u)$ be an outgoing non-timelike vector, where $s^v>0$ and $s^u\le 0$ are satisfied.
The variation formulae (\ref{m_v}) and (\ref{m_u}), and the dominant energy condition (\ref{dec}) yield $m_{,v} \ge0$ and $m_{,u} \le 0$ on an untrapped surface.
Thus we obtain $\mathscr L_s m= s^vm_{,v}+s^um_{,u} \ge 0$ on an untrapped surface.
The proof is similar for an ingoing non-timelike direction. 
\qed

\bigskip

Next we move on to the proof of positivity.
The point where $r=0$ is called {\it center} if it
defines the boundary of $(M^2, g_{ab})$.
 A central point is called {\it regular} if 
\begin{equation}
\label{r-center}
\frac{2}{(n-2)^2}e^f r^2\theta _+\theta _- +k \simeq Cr^2
\end{equation}  
holds around the center and {\it singular} otherwise, 
where a constant $C$ is assumed to be non-zero.

\begin{lm}
\label{lm:center-ex}
If $-{\tilde \Lambda}+C+{\tilde \alpha}C^2 >(<)0$ holds, 
then $m$ is positive (negative) around the regular center.
\end{lm}
\noindent
{\it Proof}. 
From Eq.~(\ref{qlm2}), we obtain
\begin{align}
\label{qlm-center}
m \simeq \frac{(n-2)V_{n-2}^k}{2\kappa_n^2}r^{n-1}
(-{\tilde \Lambda}+C+{\tilde \alpha}C^2) 
\end{align}  
around the regular center.
\qed

\begin{lm}
\label{lm:center0}
If the regular center is surrounded by untrapped surfaces and the dominant energy condition holds, then $-{\tilde \Lambda}+C+{\tilde \alpha}C^2 \ge0$ is satisfied and consequently $m$ is non-negative around the regular center.
\end{lm}
\noindent
{\it Proof}. 
From Eq.~(\ref{qlm-center}), we have
\begin{align}
m_{,v} &\simeq \frac{(n-1)V_{n-2}^k}{2\kappa_n^2}
r^{n-1}\theta_{+}(-{\tilde \Lambda}+C+{\tilde \alpha}C^2),
 \label{m_v-c} \\
m_{,u} &\simeq \frac{(n-1)V_{n-2}^k}
{2\kappa_n^2}r^{n-1}\theta_{-}
(-{\tilde \Lambda}+C+{\tilde \alpha}C^2) \label{m_u-c}
\end{align}  
around the regular center.
By Eqs.~(\ref{m_v-c}) and (\ref{m_u-c}) and Proposition~\ref{th:monotonicity}, 
if the regular center is surrounded by untrapped surfaces and the dominant energy condition holds, the inequality 
$-{\tilde \Lambda}+C+{\tilde \alpha}C^2 \ge 0$ is satisfied.
Then, by Lemma~\ref{lm:center-ex}, $m$ is non-negative around the center.
\qed

\begin{Prop}
\label{th:positivity}
({\it Positivity.}) 
If the dominant energy condition holds on an untrapped spacelike hypersurface with a regular center, 
then $m\ge 0$ holds there.
\end{Prop}
\noindent
{\it Proof}.
The proposition follows from Proposition~\ref{th:monotonicity} 
and Lemma~\ref{lm:center0}.
\qed

\bigskip

In Proposition~\ref{th:positivity}, it is assumed that a regular center is surrounded by untrapped surfaces.
By Eq.~(\ref{r-center}), a regular center is surrounded independent of $C$ by untrapped and trapped surfaces for $k=1$ and $-1$, respectively. Therefore, the positivity of $m$ is shown for $k=1$, but not for $k=-1$ because the assumption cannot be satisfied for $k=-1$.
In the case of $k=0$, on the other hand, the assumption gives a constraint on the value of $C$.

\begin{lm}
\label{lm:center-gr}
Suppose the dominant energy condition in the case of $k=0$ in general relativity.
Then, a regular center cannot be surrounded by untrapped surfaces for $\Lambda \ge 0$.
On the other hand, if a regular center is surrounded by untrapped surfaces for $\Lambda<0$, $C$ satisfies ${\tilde\Lambda} \le C <0$.
\end{lm}
\noindent
{\it Proof}. 
Suppose the dominant energy condition and the regular center surrounded by untrapped surfaces.
Then, $C$ is negative by Eq.~(\ref{r-center}), 
while $C \ge {\tilde\Lambda}$ holds by Lemma~\ref{lm:center0}.
Therefore, $C$ satisfies ${\tilde\Lambda} \le C <0$ if $\Lambda<0$, while $\Lambda \ge 0$ gives a contradiction.
\qed

\bigskip

Thus, in the case of $k=0$ in general relativity, the regular center 
surrounded by untrapped surfaces under the dominant energy condition 
was shown to be possible only in the presence of a negative cosmological constant. 
In Einstein-Gauss-Bonnet gravity, the constraint on the value of $C$ is more complicated.

\begin{lm}
\label{lm:center-gb}
Suppose the dominant energy condition in the case of $k=0$ in Einstein-Gauss-Bonnet gravity.
Then, if a regular center is surrounded by untrapped surfaces, $C$ satisfies $C <0$ if $\tilde \Lambda \le -1/(4{\tilde\alpha})$, $C< C_-$ or $C_+<C<0$ if $-1/(4{\tilde\alpha})<\tilde \Lambda <  0$, and $C \le C_-$ if ${\tilde \Lambda} \ge 0$, where 
$C_+:=(-1+\sqrt{1+4{\tilde\alpha}{\tilde\Lambda}})/(2{\tilde\alpha})$ 
and $C_-:=(-1-\sqrt{1+4{\tilde\alpha}{\tilde\Lambda}})/(2{\tilde\alpha})$.
\end{lm}
\noindent
{\it Proof}. 
Suppose the dominant energy condition and the regular center surrounded by untrapped surfaces.
Then, $C$ is negative by Eq.~(\ref{r-center}), 
while $-{\tilde \Lambda}+C+{\tilde \alpha}C^2 \ge 0$ holds by Lemma~\ref{lm:center0}.
The latter inequality is satisfied for any $C$ if $1+4{\tilde\alpha}{\tilde\Lambda} \le 0$.
If $1+4{\tilde\alpha}{\tilde\Lambda} > 0$, it is satisfied for $C$ satisfying $C \le C_-<0$ or $C \ge C_+$, where $C_+>(<)0$ holds for positive (negative) $\Lambda$ and $C_+=0$ holds for $\Lambda=0$.
\qed

\bigskip

In the positivity proof of the Misner-Sharp mass ($n=4$, $k=1$, and $\Lambda=0$) in~\cite{hayward1996}, it is claimed that Proposition~\ref{th:positivity} follows immediately from Proposition~\ref{th:monotonicity} together with the fact that a regular center is surrounded by untrapped surfaces by Eq.~(\ref{r-center}).
However, because the sign of $m$ around the regular center depends on the value of $C$ as seen in Lemma~\ref{lm:center-ex}, the positivity of $m$ around the regular center seems to be nontrivial, which requires Lemma~\ref{lm:center0} for completion of the proof.

As mentioned above, the proof of Proposition~\ref{th:positivity} does not work for $k=-1$ and for $k=0$ depending on $C$ in Eq.~(\ref{r-center}).
However, under the special relation (\ref{DCgravity}) between the coupling constants, with which our theory reduces to Chern-Simons gravity for $n=5$~\cite{Banados:1993ur,zanelli2005}, the positivity of $m$ is shown for any $k$ without assumptions in Proposition~\ref{th:positivity}.

\begin{Prop}
\label{th:positivity-sp}
({\it Positivity with $1+4{\tilde\alpha}{\tilde\Lambda}=0$.}) 
If $1+4{\tilde\alpha}{\tilde\Lambda}=0$, then $m \ge 0$ holds.
\end{Prop}
\noindent
{\it Proof}.
For $1+4{\tilde\alpha}{\tilde\Lambda}=0$, Eq.~(\ref{qlm}) gives 
\begin{align}
m = \frac{(n-2)V_{n-2}^k}{8{\tilde \alpha}\kappa_n^2}r^{n-5}
\biggl\{r^2
+2{\tilde \alpha}[k-(D r)^2]\biggl\}^2 \ge 0.
\end{align}  
\qed

\bigskip

In the asymptotically AdS case under the special relation (\ref{DCgravity}), moreover, the following result is obtained.

\begin{Prop}
\label{th:positivity-sp2}
({\it Vanishing in asymptotically AdS spacetime with $1+4{\tilde\alpha}{\tilde\Lambda}=0$.}) 
Suppose $1+4{\tilde\alpha}{\tilde\Lambda}=0$ and the dominant energy condition in an $n$-dimensional asymptotically AdS spacetime.
Then, $m=0$ holds on an untrapped spacelike hypersurface.
\end{Prop}
\noindent
{\it Proof}.
For $1+4{\tilde\alpha}{\tilde\Lambda}=0$, we have $m=0$ at infinity by Proposition~\ref{th:asymptotics2}.
Thus, by Propositions~\ref{th:monotonicity} and \ref{th:positivity-sp}, $m=0$ holds on an untrapped spacelike hypersurface.
\qed

\bigskip

Here we note that, although the metric in the generalized Boulware-Deser-Wheeler solution~(\ref{BDW}) for $n \ge 6$ with $k=1$ and $1+4{\tilde\alpha}{\tilde\Lambda}=0$ approaches AdS at infinity for an arbitrary positive constant $M$ and $M$ coincides with our quasi-local mass, it does not conflict with Proposition~\ref{th:positivity-sp2}.
This is because that spacetime is not asymptotically AdS in the sense that the fall-off condition (\ref{bc}) does not hold.

The positivity property of the quasi-local mass has a physical interpretation whereby 
under the stated circumstances the sum of the matter energy 
and the gravitational potential energy cannot be negative.
This is not obvious even when an energy condition on matter is assumed
since gravitational potential energy tends to be negative~\cite{hayward1996}.
The results of this section are summarized in Table~\ref{table:qlm}.

\begin{widetext}
\begin{center}
\begin{table}[h]
\caption{\label{table:qlm} Properties of the quasi-local mass.
For $k=-1$, the assumption in Proposition~\ref{th:positivity} for positivity cannot be satisfied, while it constrains the value of $C$ for $k=0$. (See Lemmas~\ref{lm:center-gr} and \ref{lm:center-gb}.) In the special case where $1+4{\tilde\alpha}{\tilde\Lambda}=0$, positivity of $m$ is shown for any $k$ without assumptions in Proposition~\ref{th:positivity}.}

\begin{tabular}{l@{\qquad}c@{\qquad}c@{\qquad}c}
\hline \hline
  & $k=1$ & $k=0$ & $k=-1$   \\\hline
Unified first law & Yes & Yes & Yes \\ \hline
Global mass & Yes & Not applicable & Not applicable \\ \hline
Monotonicity & Yes & Yes & Yes \\ \hline
Positivity & Yes & See the caption & See the caption \\
\hline \hline
\end{tabular}
\end{table} 
\end{center}
\end{widetext}

\section{Summary and discussion}
\label{sec:summary}

A quasi-local mass characterizes spacetime geometry quasi-locally and represents the energy enclosing a spatial surface. 
In the present paper, we have analyzed properties of the generalization of the Misner-Sharp quasi-local mass in a higher-dimensional spacetime having symmetries corresponding to the isometries of an $(n-2)$-dimensional maximally symmetric space in Einstein-Gauss-Bonnet gravity. 
Our quasi-local mass is defined in a purely geometrical way and reduces to the Misner-Sharp mass in the four-dimensional spherically symmetric case without a cosmological constant.

It was shown that our quasi-local mass (\ref{qlm}) possesses properties similar to those of the Misner-Sharp mass.
Our quasi-local mass coincides with a charge associated with a locally conserved current constructed from the generalized Kodama vector and satisfies the unified first law, which states that the change of the quasi-local mass is complemented by the energy inflow and the external work. 
This should be one of the touchstones of the quasi-local mass. 
We also classified all the vacuum solutions by utilizing the generalized Kodama vector.

The quasi-local mass satisfies the simple variation formulae (\ref{m_v}) and (\ref{m_u}), which are the same as those in general relativity. 
As a result, they allow us to prove the monotonicity and positivity of the quasi-local mass in a similar manner to the general relativistic case.
Under the dominant energy condition, monotonicity on an untrapped surface and positivity on an untrapped spacelike hypersurface with a regular center were shown to hold.
However, we also showed that the assumptions in the proof of positivity are not realized for $k=-1$ and for $k=0$ with a non-negative cosmological constant in general relativity.
In contrast, under a special relation (\ref{DCgravity}) between coupling constants, positivity of the quasi-local mass is shown for any $k$ without assumptions above.

It was shown that our quasi-local mass approaches the higher-dimensional global mass at (spacelike) infinity in the asymptotically flat or AdS spacetime.
In the asymptotically flat case, it approaches the higher-dimensional ADM mass at spacelike infinity, while it does the Deser-Tekin and Padilla mass at infinity in the asymptotically AdS case.
On the other hand, we have not argued the asymptotic behavior of the quasi-local mass at null infinity.
The Misner-Sharp mass approaches the Bondi mass at null infinity in the vacuum case~\cite{hayward1996}.
This asymptotic property is one of the criteria for the well-posedness of a quasi-local mass.
It is tempting to hope that our quasi-local mass should be asymptotic to the higher-dimensional Bondi mass in that limit.
However, as mentioned in subsection~\ref{subsec:asymp}, this is indeed the case at least in even dimensions~\cite{hollands2005,hollands2004}. 
The absence of a stable conformal null infinity forbids us from defining the Bondi-like radiation energy for odd-dimensional spacetimes in terms of the conformal completion technique. 
We have at present no alternative way of dealing with the radiation energy but to make use of conformal infinity.
The meaning of the radiation energy in the asymptotically flat case remains open in odd dimensions.

All above results support the interpretation of $m$ defined by Eq.~(\ref{qlm}) as a well-posed quasi-local mass at least in the spherically symmetric case. 
One of the main applications of the quasi-local mass is to the black-hole dynamics. 
In dynamical spacetime, a black hole can be locally defined by a future outer trapping horizon~\cite{hayward1996}.
Then, the quasi-local mass can be used to evaluate the mass of such a dynamical black hole.
Actually, we can read off the dynamical black-hole entropy by rewriting the unified first law.
This issue will be reported in a subsequent paper~\cite{nm2007}.

\bigskip

We conclude this paper by speculation about further generalization of the quasi-local mass in Lovelock gravity.
Einstein-Gauss-Bonnet gravity as well as general relativity give rise to the quasi-linear second-order field equations and are classes of Lovelock gravity~\cite{lovelock}. 
Lovelock gravity exhibits some remarkable properties. 
When we write the field equations as 
${\ma G}_{\mu \nu }=\kappa ^2T_{\mu \nu }$, 
(1) ${\ma G}_{\mu \nu }$ is symmetric in its indices,
(2) ${\ma G}_{\mu \nu }$ contains up to the second derivative of the metric,
(3) $\nabla _\nu {\ma G}^{\mu\nu} \equiv 0$, and 
(4) ${\ma G}_{\mu \nu }$ is linear in the second derivative of the metric.
In four dimensions, the fourth condition is derived by other three.
The Lovelock Lagrangian comprises the dimensionally extended
Euler densities.
In $n$-dimensional spacetimes, up to [$n/2$]-curvature terms
appear in the field equations, where [$x$] denotes the integer part of $x$. 
But in even dimensions, the last ($(n/2)$-th) term  does not contribute to field equations because it becomes a topological invariant. 
Then, a natural question arises, whether a similar quasi-local mass 
can be defined in Lovelock gravity?

The action for Lovelock gravity is given by 
\begin{equation} 
\label{actionL}
S=\frac{1}{2\kappa_n^2}\int \D ^nx\sqrt{-g}
\sum_{i=0}^{[n/2]}\alpha_i{\ma L}_{(i)}+S_{\rm matter},
\end{equation}
where ${\ma L}_{(i)}$ is the $i$-th order Lovelock Lagrangian, which is an $i$-th polynomial in Riemann curvature and its contractions, 
and we identify ${\ma L}_{(0)} := 1$, 
${\ma L}_{(1)} := R$, ${\ma L}_{(2)} := L_{\rm GB}$ and so on~\cite{lovelock}.
$\alpha_i$ is a coupling constant with dimension 
$({\rm length})^{2(i-1)}$ such as 
$\alpha_0 := -2\Lambda$, $\alpha_1 := 1$, and $\alpha_2 := \alpha$.
The gravitational equation following from this action is given by 
\begin{equation} 
{\ma G}_{\mu\nu} := \sum_{i=0}\alpha_{i}{G}^{(i)}_{\mu\nu}
=\kappa_n^2 {T}_{\mu\nu}, \label{beqL}
\end{equation} 
where the tensor ${G}^{(i)}_{\mu\nu}$ is given from ${\ma L}_{(i)}$ 
such as ${G}^{(0)}_{\mu\nu} := -(1/2)g_{\mu\nu}$, 
${G}^{(1)}_{\mu\nu} :=G_{\mu\nu}$, 
and ${G}^{(2)}_{\mu\nu} := H_{\mu\nu}$.

We propose the generalized Misner-Sharp quasi-local mass in Lovelock gravity: 
\begin{equation}
\label{qlmL}
m_{\rm L} := \frac{V_{n-2}^k}{2\kappa_n^2}\sum_{i=0}^{[n/2]}
\frac{\alpha_i (n-2)!}{(n-1-2i)!}r^{n-1-2i}[k-(Dr)^2]^i.
\end{equation}
$m_{\rm L}$ would approach to the higher-dimensional ADM mass at spacelike infinity in an asymptotically flat spacetime because higher-order curvature terms fall off sufficiently rapidly.

We envisage that the unified first law continues to be
valid in Lovelock gravity.  
\begin{Conj}
({\it Unified first law.}) 
\label{conjecture0}
The unified first law (\ref{1stlaw1}) holds in Lovelock 
gravity by replacing $m$ by $m_{\rm L}$.
\end{Conj}
Since the unified first law gives us a clear physical interpretation, the validity of above conjecture will enhance the reliability of the quasi-local mass.

Conjecture~\ref{conjecture0} directly implies that the variation formulae (\ref{m_v}) and (\ref{m_u}) hold in Lovelock gravity by replacing $m$ by $m_{\rm L}$.
As seen in Propositions~\ref{th:monotonicity} and \ref{th:positivity}, the monotonicity and positivity of the
quasi-local mass are easily shown by these variation formulae under the dominant energy condition.
Thus, this conjecture implies that they also holds in Lovelock gravity.
Conjecture~\ref{conjecture0} also implies that the generalized Misner-Sharp mass formalism in Lovelock gravity would be available in the system with a perfect fluid satisfying $p \ne -\rho$, which is obtained by replacing $m$ by $m_{\rm L}$ in Eqs.~(2.15)--(2.20) in~\cite{maeda2006b}.

We speculate that the following local conservation laws would hold in Lovelock gravity.
\begin{Conj}
({\it Local conservation law.}) 
For the generalized Kodama vector $K^\mu$, 
\begin{eqnarray}
{G}^{(i)}_{\mu\nu}\nabla ^\nu K^{\mu} \equiv 0
\end{eqnarray}  
holds, so that 
\begin{eqnarray}
J^{(i)\mu} := {G^{(i)\mu}}_{\nu}K^{\nu}
\end{eqnarray}  
is divergence-free for each $i$.
Then, $\mathscr L_J m_{\rm L}=0$ holds and $m_{\rm L}$ is given as 
\begin{align}
m_{\rm L} &= \int _\Sigma J^\mu \D \Sigma _\mu,\\
J^\mu &:= -\frac{1}{\kappa _n^2}\sum_{i=0}^{[n/2]}\alpha_i J^{(i)\mu},
\end{align}
where the integration is done over some spatial volume $\Sigma$ with a boundary, as shown in section~\ref{sec:Kodamavector}.
\end{Conj}

Properness of above two conjectures give a possibility to treat any class of Lovelock gravity in a unified manner.
They will be quite helpful to give us much deeper insights into Lovelock gravity.


\section*{Acknowledgments}
The authors would like to thank L\'aszl\'o B. Szabados, Tomohiro 
Harada and Kei-ichi Maeda for useful discussions.
We are also thankful to Matthias Blau and an anonymous referee for valuable comments. 
HM would like to thank the Department of Physics, National Central 
University, warmly for hospitality facilitating this work.  
HM was supported by Grant No. 1071125 from FONDECYT (Chile) and the 
Grant-in-Aid for Young Scientists (B), 18740162, from the Scientific 
Research Fund of the Ministry of Education, Culture, Sports, Science and 
Technology (MEXT) of Japan. CECS is funded in part by an institutional 
grant from Millennium Science Initiative, Chile, and the generous 
support to CECS from Empresas CMPC is gratefully acknowledged.
MN was partially supported by JSPS.

\appendix

\section{Curvature tensors}
\label{sec:curvature}

The non-vanishing components of the Levi-Civit\'a connections are
\begin{align}
\begin{aligned}
{\Gamma ^a}_{bc}&={}^{(2)}{\Gamma ^a}_{bc }(y),\quad 
{\Gamma ^i}_{ij}={\hat{\Gamma} ^i}_{~jk}(z), \\
{\Gamma ^a}_{ij}&=-r (D^a r) \gamma _{ij},\quad 
{\Gamma ^i}_{ja}=\frac{D_a r}{r}{\delta ^i}_j, 
\end{aligned}
\end{align}
where the superscript (2) denotes the two-dimensional quantity,
and $D_a$ is the two-dimensional linear connection compatible with
$g_{ab}$. ${\hat \Gamma ^i}_{~jk}$ is the Levi-Civit\'a connection
associated with $\gamma _{ij}$.
The Riemann tensor is given by
\begin{align}
{R^a}_{bcd}&={}^{(2)}{R^a}_{bcd},\nonumber \\
{R^a}_{ibj}&=-r(D^a D_b r)\gamma _{ij},
\label{eq:Riemann}\\
{R^i}_{jkl}&=[k-(Dr)^2]({\delta ^i}_k\gamma _{jl}
-{\delta ^i}_l\gamma _{jk}), \nonumber 
\end{align}
The Ricci tensor and the Ricci scalar are given by 
\begin{align}
R_{ab}&={}^{(2)}R_{ab}-(n-2)\frac{D_aD_br}{r}, \nonumber \\
R_{ij}&=\left\{-r D^2r+(n-3)[k-(Dr)^2]\right\}
\gamma _{ij}, \label{eq:Ricci} \\
R&={}^{(2)}R-2(n-2)\frac{D^2r}{r}
+(n-2)(n-3)\frac{k-(Dr)^2}{r^2}. \nonumber
\end{align}
The Weyl tensor is simplified to
\begin{align}
C_{abcd}&=\frac{n-3}{n-1} Wg_{a[c}g_{d]b},\nonumber \\
C_{aibj}&=-\frac{n-3}{2(n-1)(n-2)}W g_{ab}r^2\gamma _{ij}, \\
C_{ijkl}&=\frac{2}{(n-1)(n-2)}
Wr^4\gamma _{i[k}\gamma_{l]j},
\nonumber
\end{align}
with 
\begin{align}
W:={}^{(2)}R+2\frac{D^2
 r}r+2\frac{k-(Dr)^2}{r^2}.
\end{align}
\begin{widetext}
Availing ourselves of the following identity
\begin{align}
\biggl(D_a D_b r-\frac 12 g_{ab}D^2 r \biggr)D^2 r 
\equiv (D_aD^c r)(D_bD_cr) -\frac 12 g_{ab}(D_cD_d r)(D^cD^d r),
\label{S=0}
\end{align}
we express the Gauss-Bonnet tensor as
\begin{align}
\begin{aligned}
H_{ab}=&\frac{2(n-2)(n-3)(n-4)}{r^3}[k-(Dr)^2]\left[
\left\{
D^2r-(n-5)\frac{[k-(Dr)^2]}{4r}
\right\}
g_{ab}- D_aD_br \right], \\
H_{ij}=& 2(n-3)(n-4)\left[
-\frac {k-(Dr)^2}2{}^{(2)}R
- (D^2r)^2 +(D_aD_br)(D^aD^br)\right. 
\\
&\left. \qquad \qquad \qquad \quad -(n-5)(n-6)
\frac{[k-(Dr)^2]^2}{4 r^2}
+(n-5)\frac{k-(Dr)^2}{r}D^2r
\right]\gamma _{ij}. 
\end{aligned}
\end{align}
The Gauss-Bonnet combination is given by
\begin{align}
L_{\rm GB}=\frac{4 (n-2) (n-3)}{r^2} 
\biggl[&\frac{k-(Dr)^2}{2 }{}^{(2)}R+(D^2 r)^2-{(D_a D_b r) (D^a D^b r)} \nonumber \\
&+(n-4)(n-5)\frac{[k-(D r)^2]^2}{4 r^2}-(n-4)
\frac{k-(Dr)^2}{r}D^2 r\biggl].
\end{align}
\end{widetext}


\end{document}